\documentclass[sigconf]{acmart}

\usepackage[show]{chato-notes}
\usepackage{marginnote}
\usepackage{minted}
\usepackage{listings}
\usepackage{xcolor}
\usepackage{flushend}
\usepackage{multicol}

\lstdefinestyle{mystyle}{
    language=Python,
    basicstyle=\ttfamily\footnotesize,
    keywordstyle=\color{blue}\bfseries,
    commentstyle=\color{gray}\itshape,
    stringstyle=\color{red},
    numberstyle=\tiny\color{gray},
    numbers=left,
    stepnumber=1,
    numbersep=5pt,
    backgroundcolor=\color{lightgray!20},
    frame=single,
    rulecolor=\color{black},
    breaklines=true,
    breakatwhitespace=true,
    captionpos=b,
    tabsize=4,
    showspaces=false,
    xleftmargin=10pt,    %
    aboveskip=5pt,      %
    belowskip=5pt,      %
    columns=flexible,    %
}
\newcommand{\cm}[1]{\textcolor{black}{#1}}
\newcommand{\ap}[1]{\textcolor{black}{#1}}
\usepackage{fontawesome5}

\usepackage{marginnote}

\newcommand{\pageenlarge}[1]{\marginnote{}\enlargethispage{#1\baselineskip}}

\copyrightyear{2025}
\acmYear{2025}
\setcopyright{rightsretained}
\acmConference[SIGIR '25]{Proceedings of the 48th International ACM SIGIR Conference on Research and Development in Information Retrieval}{July 13--18, 2025}{Padua, Italy}
\acmBooktitle{Proceedings of the 48th International ACM SIGIR Conference on Research and Development in Information Retrieval (SIGIR '25), July 13--18, 2025, Padua, Italy}\acmDOI{10.1145/3726302.3730150}
\acmISBN{979-8-4007-1592-1/2025/07}

\usepackage{amsmath}
\usepackage{natbib}
\usepackage{amsfonts}
\usepackage{graphicx}
\usepackage{subcaption}

\usepackage{hhline}
\usepackage{hyperref}
\usepackage{enumerate}
\usepackage{multirow}
\usepackage{array}
\usepackage[para]{footmisc}
\usepackage{marginnote}

\newcommand{\jy}[1]{\textcolor{black}{#1}}%

 \author{Craig Macdonald}
\affiliation{\institution{University of Glasgow}
\city{Glasgow}
\country{United Kingdom}}

 \author{Jinyuan Fang}
\affiliation{\institution{University of Glasgow}
\city{Glasgow}
\country{United Kingdom}}

 \author{Andrew Parry}
\affiliation{\institution{University of Glasgow}
\city{Glasgow}
\country{United Kingdom}}

 \author{Zaiqiao Meng}
\affiliation{\institution{University of Glasgow}
\city{Glasgow}
\country{United Kingdom}}

\begin{document}

\title[Constructing and Evaluating Declarative RAG Pipelines in PyTerrier]{Constructing and Evaluating Declarative RAG Pipelines \\ in PyTerrier}

\begin{abstract}
Search engines often follow a pipeline architecture, where complex but effective reranking components are used to refine the results of an initial retrieval. Retrieval augmented generation (RAG) is an exciting application of the pipeline architecture, where the final component generates a coherent answer for the users from the retrieved documents. In this demo paper, we describe how such RAG pipelines can be formulated in the declarative PyTerrier architecture, and the advantages of doing so. Our PyTerrier-RAG extension for PyTerrier provides easy access to standard RAG datasets and evaluation measures, state-of-the-art LLM readers, and using PyTerrier's unique operator notation, easy-to-build pipelines. We demonstrate the succinctness of indexing and RAG pipelines on standard datasets (including Natural Questions) and how to build on the larger PyTerrier ecosystem with state-of-the-art sparse, learned-sparse, and dense retrievers, and other neural rankers.
\end{abstract}




\maketitle

 \section{Introduction}
\pageenlarge{1} \looseness -1 Information retrieval is increasingly concerned with complex architectures, beyond single-step sparse retrieval: for instance, learning-to-rank brought multi-stage architectures where candidate documents had extract features calculated, before being {\em re-ranked} by learned models~\cite{Liu09ftir}; more recently neural rerankers, such as those based on language models have become popular~\cite{lin2021pretrainedtransformerstextranking}; \cm{Similarly} for dense retrievers, different pipeline stages can be conceived: query encoding, retrieval (including exact or approximate neighbourhood retrievers such as HNSW~\cite{10.1109/TPAMI.2018.2889473} or FAISS~\cite{FAISS}), or even pseudo-relevance feedback~\cite{wang2021pseudo,10.1145/3570724}.

PyTerrier\footnote{\faGithubSquare{}~\url{ https://github.com/terrier-org/pyterrier}}~\cite{pyterrier_ictir,macdonald:cikm2021-pyterrier} is a research platform that embraces multi-stage architectures for information retrieval research, by allowing indexing and retrieval pipelines to be formulated as expressions in Python: a rich ecosystem of transformations (indexers, retrievers, rerankers, features extractors) can be combined using intuitive operators into coherent pipelines. Each such transformation is applied on a common data model (i.e. data types or {\em relations} with expected attributes). The resulting pipelines are declarative in nature, in that the pipeline is formulated without considering the queries on which it will be applied - the resulting code is therefore declarative in nature and easily readable. %
Finally, pipelines can be evaluated and compared in a declarative experiment API, allowing the easy comparison of different retrieval systems for different query datasets, evaluation measures etc. Such evaluation avoids creating intermediate files (which are implementation by-products), as would have been required for traditional IR evaluation tools such as {\tt trec\_eval}.

On the other hand, while large language models (LLMs) have shown excellent memorisation able to often generate \cm{accurate answers to questions}, they can hallucinate; this has given rise to Retrieval Augmented Generation (RAG)~\cite{lewis2020retrieval} as an increasingly popular multi-stage architecture, where the LLMs are provided the retrieved document as context in order to improve the quality of the generated answer. \jy{So-called {\em sequential} RAG models often employ a \textit{retriever-reader} architecture, where a retriever first retrieves documents relevant to the query and a reader then generates an output based on the retrieved documents~\cite{gao2023retrieval}. By grounding LLM outputs in retrieved documents, RAG models have demonstrated remarkable performance in a variety of tasks, such as question answering (QA)~\cite{lin2024ra}, fact checking~\cite{lewis2020retrieval} and dialogue generation~\cite{wang2024unims}. However, sequential RAG models often struggle with tasks that require multi-step reasoning, such as multi-hop QA~\cite{trivedi2023interleaving}. This limitation arises because single-step retrieval may fail to retrieve all the relevant information, leading to knowledge gaps in the reasoning process~\cite{shao2023enhancing}. To address this issue, recent research has proposed {\em iterative} or adaptive RAG models, which perform multiple rounds of retrieval and reasoning to progressively gather the necessary information~\cite{trivedi2023interleaving,asai2024self,su2024dragin} and determine the answer~\cite{fang2024reano,fang2024trace}. By dynamically retrieving additional information as needed, these approaches enhance RAG models' ability to handle complex reasoning tasks.}

\pageenlarge{2}  \looseness -1 In particular, to \cm{benefit} RAG researchers, we bring PyTerrier-RAG, which supports (i) QA datasets (e.g. Natural Questions, NQ~\cite{kwiatkowski2019natural}) - including providing pre-built indices, (ii) evaluation measures for QA (e.g. EM\%, F1~\cite{chen2017reading,yang2018hotpotqa}), and access to a number of recent answer generation approaches. This is supplemented by standard classes allowing common styles of RAG approaches to be easily implemented, including iterative approaches characterised by IRCoT~\cite{trivedi2023interleaving}. Indeed, by implementing these in PyTerrier, RAG experimentation with other retrieval components is as easy as renaming one variable to another. Underlying all of this, an extended PyTerrier data model for RAG, and the intuitive PyTerrier operators allow pipelines to be constructed easily. PyTerrier-RAG's source code and \cm{interactive Google Colab notebooks} can be found at \faGithubSquare{}~\url{https://github.com/terrierteam/pyterrier\_rag}.

\looseness -1 The remainder of this paper is structured as follows: We position  PyTerrier-RAG with respect to other existing RAG frameworks in Section~2; In Section~\ref{sec:pyterrier}, we provide an overview of the necessary preliminaries of PyTerrier; Section~4 provides an overview of PyTerrier-RAG, describing advances in terms of data model, pipeline components, datasets and evaluation. We provide a full worked example in Section~5, and provide concluding remarks in Section~6.

\section{Related Work}

\looseness -1 Among a number of commercial offerings, there are also several existing research-oriented RAG platforms that focus on different aspects of retrieval-augmented generation. For example, DSPy~\cite{khattab2023dspy} is a framework designed for programming LLMs. It allows users to create compositional modules to build complex AI systems, including RAG pipelines and agentic applications~\cite{yao2023react}. However, DSPy does not offer pre-built RAG pipelines, requiring users to implement their own pipelines. In contrast, our framework offers both pre-processed datasets and pre-implemented RAG pipelines to enable succinct RAG experimentation. 
Recently, \citet{rau2024bergen} introduced BERGEN, a benchmarking RAG system with a focus on question answering. BERGEN implements different retrievers and readers, enabling a systematic investigation of different components within a consistent setting. However, BERGEN is limited to the sequential RAG pipeline, consisting of a retriever followed by a reader. In contrast, our framework supports not only sequential pipelines but also more advanced RAG architectures, such as iterative RAG pipelines. Another platform, RAGChecker~\cite{ru2024ragchecker}, is tailored to the evaluation of RAG systems. It offers an automatic evaluation framework designed to assess each component of the RAG pipelines. 

\pageenlarge{2}\looseness -1 In addition, among the existing RAG research platforms, Flash\-RAG~\cite{jin2024flashrag} is the most similar to our framework. It provides a comprehensive environment for developing and evaluating RAG pipelines, offering pre-built datasets, retrieval models, and different readers. However, there are key differences between the two platforms. 
One key difference is that FlashRAG relies on configuration files for setup and management, whereas our framework eliminates the need for such files. This design choice in PyTerrier-RAG allows users to configure and modify their RAG pipelines directly within their code without modifying external configuration files - ideal for notebook-based agile research. Additionally, PyTerrier-RAG adopts a declarative style, allowing users to easily compose and modify different RAG pipelines in a plug-and-play manner, thereby enhancing flexibility and usability. 
Another key difference is that FlashRAG offers a limited set of retrievers and rerankers. For example, it does not support multi-vector dense retrieval models such as ColBERT~\cite{khattab2020colbert} or learned sparse retrieval models such as SPLADE~\cite{splade}, which are two families of strong retrieval models from the literature. In contrast, our framework is built upon the PyTerrier \cm{platform}, enabling access to a large ecosystem of retrievers and rerankers. 

\section{PyTerrier Preliminaries}\label{sec:pyterrier}

\looseness -1 PyTerrier operates on relations with known primary keys and optional attributes. For instance, queries are represented by the $Q$ type with schema $Q(qid, query, ...)$; and documents by the $D$ type with schema $D(docno, text, ...$). Retrieved documents, $R$, are defined as a relationship type between $Q$ and $D$, i.e. $R \subset Q \times D$, \cm{with schema} $R(qid, docno, score, rank)$; similarly relevant assessments \cm{are} also a subset of $Q \times D$, i.e. $RA(qid, docno, label)$. All these relations can be instantiated in Python as Pandas Dataframes, or more simply iterable lists of dictionaries with the required and optional attributes.

Next, each component of an indexing or retrieval pipelines forms a transformation from one relation type to one or another type. We can characterise different families of transformation that are applied in indexing \& retrieval:
\begin{itemize}
\item {\em Retrieval}: $Q \to R$
\item {\em Reranking}: $R \to R$
\item {\em Query rewriting}: $Q \to Q$
\item {\em Document expansion}: $D \to D$
\item {\em Pseudo-relevance feedback}: $R \to Q$
\item {\em Indexing}: $D \to \emptyset$ (a terminal operation)
\end{itemize}

For instance, one can create a retriever on a given index, which when provided with a set of one or more queries, returns 0 or more retrieved documents for each query. We call such components {\em transformers} as they transform from one datatype to another.

\pageenlarge{2}\looseness -1 Typically, indexing or retrieval pipelines are combinations of several such transformer components. An imperative use would involve getting results from one transformer before inputting them into another. Instead, PyTerrier suggests a declarative manner, in which pipelines are constructed before applying them to unseen input data.\footnote{Advantages include, for instance, deterring bad practices such as excessive preprocessing of queries that makes testing on unseen queries difficult to achieve.} To aid in this, several operators are defined on transformers (implemented through the use of operator overloading within Python). For example, a hybrid retriever that linearly combines the scores of two different retrievers can be expressed as {\tt ret1 + ret2}. The result of such an expression is itself a transformer. The most frequently used operator is $\gg$ (also called `then'), and analogous to the $|$ operator in Unix shells, is defined as follows {\tt a $\gg$ b}: a is applied on its input, and the output of a is then passed as input to b. Salient PyTerrier operators and their meanings are summarised in Table~\ref{tab:operators}. Each operator has relational algebra semantics, as explored in~\cite{pyterrier_ictir}.

\begin{table}[tb!]
    \centering
     \caption{Exemplar PyTerrier operators for combining transformers.}    \label{tab:operators}
    \begin{tabular}{ccp{5cm}}
    \toprule
     Op.\ & Name & Description \\
    \midrule
        \texttt{>}\texttt{>} & \textit{then} & Pass the output from one transformer to the next transformer\\
        \texttt{+} & \textit{linear combine} & Sum the query-document scores of the two retrieved results lists \\
        \texttt{|} & \textit{set union} & Make the set union of documents from the two retrieved results lists \\
        \texttt{\%} & \textit{rank cutoff} & Shorten a retrieved results list to the first $K$ elements \\
    \bottomrule
    \end{tabular}\vspace{-\baselineskip}
\end{table}

A key advantage of the PyTerrier ecosystem is the richness of its ecosystem of plugins - instead of bloating the core platform with additional functionality, our preference is that we, and others, release plugins that can be maintained separately, taking advantage of the decomposable of indexing and retrieval pipelines and the extensible data model. For instance, a dense retrieval pipeline may be expressed as {\tt query\_encoder >{\tt}> retriever}, with the $Q$-typed  output of the {\tt query\_encoder} being augmented with an additional {\em query\_emb} column. Table~\ref{tab:plugins} provides an overview of many PyTerrier plugins, and gives sample indexing and/or retrieval pipelines. 

\begin{table*}
\caption{Examples of PyTerrier plugins, with links to the corresponding GitHub repositories and sample pipelines.}\label{tab:plugins}
\resizebox{\linewidth}{!}{
\begin{tabular}{l|l|w{l}{8.5cm}}
\toprule
Plugin & Functionality & Sample Pipeline \\  
\midrule
\faGithubSquare{}~\href{https://github.com/seanmacavaney/pyterrier-anserini}{pyterrier-anserini} & Indexing \& Retrieval using Anserini~\cite{10.1145/3077136.3080721} & \footnotesize {\tt AnseriniIndex.from\_hf('macavaney/msmarco-passage.anserini').bm25()} \\
\faGithubSquare{}~\href{https://github.com/terrierteam/pyterrier_pisa}{pyterrier-pisa} & Fast Indexing \& Retrieval using PISA~\cite{mallia2019pisa,10.1145/3477495.3531656} & \footnotesize {\tt PisaIndex("./cord19-pisa").bm25() }\\
\faGithubSquare{}~\href{https://github.com/terrierteam/pyterrier_t5}{pyterrier-t5} & monoT5 \& duoT5 reranking~\cite{pradeep2021expandomonoduodesignpatterntext} & \footnotesize {\tt bm25 $>>$ MonoT5() \% 10 $>>$ DuoT5()} \\
\faGithubSquare{}~\href{https://github.com/cmacdonald/pyt_splade}{pyterrier-splade} & SPLADE indexing and retrieval & %
\footnotesize {\tt splade\_ret = splade $>>$ pt.terrier.Retriever(index, wmodel="Tf") } \\
\faGithubSquare{}~\href{https://github.com/terrierteam/pyterrier_doc2query}{pyterrier-doc2query} & Doc2Query($-$$-$)~\cite{pradeep2021expandomonoduodesignpatterntext,minusminus} document expansion & \footnotesize {\tt index\_pipe = Doc2Query() $>>$ pt.terrier.IterDictIndexer()} \\
\faGithubSquare{}~\href{https://github.com/emory-irlab/pyterrier_genrank}{pyterrier-genrank} &  Listwise generative rerankers~\cite{dhole2024pyterriergenrankpyterrierpluginreranking} & \footnotesize {\tt bm25 $>>$ LLMReRanker("castorini/rank\_vicuna\_7b\_v1")} \\
\faGithubSquare{}~\href{https://github.com/terrierteam/pyterrier_dr/blob/main/pyterrier_dr/flex/core.py}{pyterrier-dr} & Dense Retrieval & \footnotesize {\tt TasB() $>>$ FlexIndex("./cord19")}  \\
\faGithubSquare{}~\href{https://github.com/terrierteam/pyterrier_colbert}{pyterrier-colbert} & ColBERT~\cite{khattab2020colbert} Dense Retrieval & \footnotesize {\tt ColBERTFactory().end\_to\_end()} \\
\bottomrule
\end{tabular}}
\end{table*}

\looseness -1 The rich PyTerrier ecosystem demonstrates the variety of different retrieval pipelines that can be used for retrieval augmented generation. In the next section, we highlight the needed additions to support experimentation within retrieval augmented generation.

\section{PyTerrier-RAG}

To better support RAG, and allow researchers working on RAG full access to the PyTerrier retrieval and reranking ecosystem, PyTerrier-RAG makes several new \cm{additions:} (i) an extended data model (Section~\ref{ssec:datamodel}); (ii)  a suite of common answer generators (often called {\em readers}, Section~\ref{ssec:readers}), and ways to translate retrieval output into context for a language-model based reader; (iii) support for iterative RAG architectures~\cite{trivedi2023interleaving}; (iv) access to standard QA datasets (Section~\ref{ssec:dataset}); and evaluation measures (Section~\ref{ssec:evaluation}).

\subsection{Data Model Extensions}\label{ssec:datamodel}
\looseness -1 PyTerrier's primary datatypes, as outlined in Section~3, include $Q$, a set of queries/questions, and $R$, a set of retrieved documents. \cm{Building on these, we add datatypes for standard RAG QA experiments:}
answers for a question/query, $A(qid, qanswer)$; gold answers $GA(qid, [ganswer])$; \cm{and, extending $Q$, $Qc$ which contains context, i.e.\ $Q_c(qid, query, qcontext)$.} 
Using these datatypes, we can define new classes of transformations between datatypes:
\begin{itemize}
    \item 0-shot answer generation: ($Q \to A$).
    \item \cm{Creation of RAG Context: $R \to Qc$ - formulates context from a retriever.}
    \item Reader: $Qc \to A$ - takes retrieved documents as context to an LLM to generate an answer.
\end{itemize}
We next describe the readers implemented within PyTerrier-RAG.

\pageenlarge{2}  \subsection{New Reader Components}\label{ssec:readers}

\looseness-1 The primary novel component of this framework is the {\tt Reader} transformer ($Qc\rightarrow A$). For a given input text, a reader provides a single answer. To allow flexibility in models, we implement both locally served and API-based language models under a backend class. A {\tt Backend} implementation need only implement a {\tt .generate()} function that accepts an iterable of prompt strings and returns an iterable of answer strings; overall, data structure processing is handled by the {\tt Reader} - e.g.` formulation as $A$ datatype.

\begin{lstlisting}
causal_backend = HuggingFaceBackend("meta-llama/Llama-3.1-8B")
seq2seq_backend = Seq2SeqLMBackend("google/flan-t5-base")
openai_backend = OpenAIBackend('gpt-4o-mini')
\end{lstlisting}

\looseness -1 Though a RAG pipeline is generally considered to be a retrieval system followed by a generative component, we further abstract the generative component to align with the declarative nature of PyTerrier. \ap{To allow flexible setups in terms of retrieval and steps for context creation, the Reader class expects only a $Qc$ datatype, with a separate component handling the collation of intermediate context. Retrieval output is aggregated by a separate {\tt Concatenator} class, i.e $R \to Qc$. So a sequential RAG pipeline is as follows:}
\begin{lstlisting}
openai_reader = Reader(backend=openai_backend)
bm25 = pt.Artifact.from_hf('pyterrier/ragwiki-terrier').bm25()
bm25_monot5 = bm25 >> MonoT5()
pipeline = bm25_monot5 >> Concatenator() >> openai_reader
\end{lstlisting}
\cm{Furthermore, within a reader, the prompt can be a string instruction or a {\tt PromptTransformer ($Q_c \rightarrow Q_c$)} to facilitate more complex instructions. The latter receives a $Qc$ dataframe with the question and context aggregated from a retriever. Internally, a prompt template format creates a single string from question and context to be passed to a {\tt Backend} within the reader.}

Finally, we also support Fusion-in-Decoder ({\tt FiDReader}) architectures~\cite{izacard:2021}; however,  due to the separate encoding of documents and fine-tuned nature of this architecture,  this class expects a ranking ($R$) as opposed to a combined context $Qc$.

\pageenlarge{2}
\subsection{Non-Sequential RAG Pipelines}\label{ssec:pipelines}

Beyond sequential RAG, several frameworks have been proposed that apply an iterative process where the output of the reader component forms part of the input of the retriever over multiple cycles. We exemplify this paradigm with an implementation of a popular iterative method, IRCoT~\cite{trivedi2023interleaving}, which, after each iteration, checks the reader's output against a user-provided exit condition (for example, checking for the presence of a stopping phrase specified by a prompt such as ``so the answer is'') as shown below:
\begin{lstlisting}
exit_condition = lambda x: "so the answer is" in x.qanswer.iloc[0]
ircot_bm25_monot5_openai = IRCOT(retriever=bm25 >> monoT5, 
                 backend=openai_backend,
                 exit_condition=exit_condition)
\end{lstlisting}
The ability to refine the retriever, e.g., by adding a monoT5 cross-encoder, is easily viewable. More recent iterative works apply reasoning on entities and relationships extracted as knowledge graphs - for instance, our own recent works in this area, REANO~\cite{fang-etal-2024-reano} and TRACE~\cite{fang-etal-2024-trace}, can be expressed as PyTerrier-RAG pipelines. For example, in TRACE, documents must have entities and corresponding knowledge relationship triples identified (offline or online, stored as extra columns in the $R$ datatype); an LLM is prompted iteratively to build a reasoning chain with knowledge triples extracted from retrieved documents to answer a question. %

\subsection{Datasets and Corpora}\label{ssec:dataset}
\begin{table}[tb]
    \centering
    \caption{Summary of datasets and corpus in PyTerrier-RAG, where ``2Wiki'' denotes ``2WikiMultihopQA''.}\label{tab:datasets}
    \resizebox{0.47\textwidth}{!}{
    \begin{tabular}{l|c|c|c|c|c}
        \toprule
        \textbf{Task} & \textbf{Dataset} & \textbf{Corpus} & \textbf{\# Train} & \textbf{\# Dev} & \textbf{\# Test} \\ 
        \midrule
        \multirow{4}{*}{\textbf{QA}}
        & Natural Questions~\cite{kwiatkowski2019natural} & Wikipedia  & \textcolor{white}{0}79,168 & \textcolor{white}{0}8,757 & \textcolor{white}{0}3,610 \\
        & TriviaQA~\cite{joshi2017triviaqa} & Wikipedia & \textcolor{white}{0}78,785 & \textcolor{white}{0}8,837 & 11,313  \\
        & WebQuestions~\cite{berant2013semantic} & Wikipedia & \textcolor{white}{00}3,778 & --- & \textcolor{white}{0}2,032 \\
        & PopQA~\cite{mallen2023when} & Wikipedia & --- & --- & 14,267 \\
        \hline
        \multirow{4}{*}{\textbf{Multi-Hop QA}}
        & HotpotQA~\cite{yang2018hotpotqa} & HotpotQA & \textcolor{white}{0}90,447 & \textcolor{white}{0}7,405 & --- \\
        & MuSiQue~\cite{trivedi2022musique} & MuSiQue &\textcolor{white}{0}19,938 & \textcolor{white}{0}2,417  & ---  \\
        & 2Wiki~\cite{ho2020constructing} & 2Wiki & \textcolor{white}{0}15,000 & 12,576 & --- \\
        & Bamboogle~\cite{press2023measuring} & Wikipedia & --- & --- & 125 \\ 
        \hline
        \multirow{1}{*}{\textbf{Dialogue Generation}}
        & WoW~\cite{dinan2019wizard} & Wikipedia & \textcolor{white}{0}63,734 & \textcolor{white}{0}3,054 & --- \\
        \hline
        \multirow{1}{*}{\textbf{Fact Checking}}
        & FEVER~\cite{thorne2018fever} & Wikipedia & 104,966 & 10,444 & --- \\
        \bottomrule
    \end{tabular}}\vspace{-1em}    
\end{table}

To facilitate succinct research with RAG pipelines, PyTerrier-RAG offers ten pre-processed benchmark datasets, covering a variety of datasets commonly used in RAG studies~\cite{gao2023retrieval}. 
Table~\ref{tab:datasets} provides a summary of these datasets, including their respective tasks, corpora and statistics. 
Specifically, our framework provides programmatic access to datasets for various retrieval-augmented tasks, including general domain QA, multi-hop QA, dialogue generation, and fact-checking. All datasets have been processed to be compatible with PyTerrier pipelines, including three key attributes: qid, query and answer. Moreover, our framework offers flexible dataset integration, allowing users to seamlessly incorporate datasets from other platforms, such as FlashRAG~\cite{jin2024flashrag}. \cm{Indexing new datasets in PyTerrier is easy - they need only be expressed as iterable dictionaries, easily obtainable from JSON files.}

\looseness -1 In addition to the benchmark datasets, our framework offers multiple retrieval corpora, such as Wikipedia, HotPotQA, 2Wiki and MuSiQue. Specifically, the Wikipedia corpus is based on the widely used December 20, 2018 Wikipedia dump released by DPR~\cite{karpukhin2020dense}, containing about 21M passages. The HotPotQA corpus, originally introduced alongside the HotPotQA dataset, consists of around 5 million documents. The 2Wiki and MuSiQue corpora are constructed from their respective contexts following the procedure outlined in~\cite{trivedi2023interleaving}, and containing about 431K and 117K documents, respectively. 

We also provide some pre-built indices for common corpora available via HuggingFace datasets, and accessed using a single line of Python code (see the first example snippet in Section~\ref{sec:example})\footnote{More details of such artefacts are provided in \cite{macavaney:sigir2025-artifact}}.

\subsection{Evaluation}\label{ssec:evaluation} \pageenlarge{2}
Evaluation for QA in PyTerrier follows the declarative nature in the platform: A {\tt pt.Experiment} is called with four essential arguments: (1) the systems being compared; (2) the queries being used for evaluation; (3) the ground truth; and (4) the evaluation measures to be calculated. QA datasets differ from classical IR datasets in that the success is typically measured in terms of the similarity to one or more accepted gold-truth answers. Hence for evaluation in PyTerrier-RAG, we use the classical {\tt pt.Experiment()} function from PyTerrier, but change (i) the type of the ground truth from $RA$ (document-level relevance assessments) to $GA$ (gold answers); and (ii) provide definitions for the standard textual overlap measures: {\em Exact Match}~\cite{chen2017reading}; {\em F1}~\cite{yang2018hotpotqa} (both based on the implementations provided in the DPR 
repository~\cite{karpukhin2020dense}); \cm{and ROUGE measures}. An example {\tt pt.Experiment()} is shown in Section~5.

\looseness -1 Another form of evaluation is the use of language models for evaluating generated answers. We provide BERTScore~\cite{bert-score}, which compares generated answers to known relevant documents~\cite{transmission}, as well as prompts for LLMs to provide ratings on answers~\cite{zheng:2023}.

{\tt pt.Experiment()} also offers other advantages such as significance testing, multiple-testing correction, ability to change the batch size. One notable recent improvement is prefix-computation~\cm{\cite{macavaney25precomputation}}. This examines the pipelines of the systems being evaluated, identifies any common prefix of the constituent pipeline components, and applies that prefix to the input topics only once. The results on the prefix can be reused for the other pipelines. For example, if an experiment was evaluating the impact on \cm{sequential RAG} answer quality of changing the number of retrieved documents from BM25, precomputation would compute the BM25 results only once.

\section{Worked Example}\label{sec:example} \pageenlarge{2}
To simplify experimentation and facilitate efficient retrieval, our framework provides a variety of pre-built indices \cm{for common datasets}, including those based on BM25 and the E5~\cite{wang2022text} dense retrieval model. These indices allow users to quickly set up retrieval pipelines without the need for extensive preprocessing or index construction. Below, we show an example experiment evaluating answer quality when comparing using the top 10 BM25 \cm{retrieved} documents versus the top 10 from E5. The output is a table of F1 and EM measurements for both pipelines -- example output is shown in the provided live notebooks. 

\begin{lstlisting}
dataset = pt.get_dataset('rag:nq')
sparse_index = pt.Artifact.from_hf('pyterrier/ragwiki-terrier')
e5_emb_index = pt.Artifact.from_hf('pyterrier/ragwiki-e5.flex')
bm25 = sparse_index.bm25(include_fields=['docno','title','text'])
e5 = E5() >> e5_emb_index.retriever()
raw_text = pt.text.get_text(sparse_index, ['title', 'text'])
fid = pyterrier_rag.readers.T5FiD("terrierteam/t5fid_base_nq")
bm25_fid = bm25 %
e5_fid = e5 %
pt.Experiment(
    [bm25_fid, e5_fid],
    dataset.get_topics('dev'),
    dataset.get_answers('dev'),
    [pyterrier_rag.measures.F1, pyterrier_rag.measures.EM])
\end{lstlisting}

\section{Conclusions}
This demonstration paper aims to summarise recent developments in a RAG plugin for PyTerrier, showing how the PyTerrier datamodel can be extended to RAG use cases, and allows state-of-the-art retrieval techniques to be easily interchanged to examine their impact upon answer quality. We believe that easily expressing different retrieval pipelines is of particular importance in RAG research, as the nature of LLMs is that consumption of information is lossy, and differ from previous expectations in the IR about ordering documents by descending likelihood of estimated relevance~\cite{transmission,10.1145/3626772.3657834}. In future work, we will continue to supplement PyTerrier-RAG with more datasets and more reader implementations.

\begin{acks}
We thank the authors of FlashRAG~\cite{jin2024flashrag} for their efforts in standardising the various QA datasets into a single HuggingFace repository. We thank our colleagues Sean MacAvaney for input in PyTerrier-RAG, and Fangzheng Tian for the BERTScore implementation.
\end{acks}

\onecolumn
\begin{multicols}{2}
\bibliographystyle{ACM-Reference-Format}
\bibliography{reference}
\end{multicols}

\end{document}